\documentclass[12pt]{article}
\usepackage{amsmath}
\usepackage{bm}
\usepackage{color}

\begin{document}
\begin{center}
\begin{large}
{\bf Upper bound on the momentum scale in noncommutative phase space of canonical type}
\end{large}
\end{center}

\centerline {Kh. P. Gnatenko \footnote{E-Mail address: khrystyna.gnatenko@gmail.com}, V. M. Tkachuk \footnote{E-Mail address: voltkachuk@gmail.com}}
\medskip
\centerline {\small \it Ivan Franko National University of Lviv, Department for Theoretical Physics,}
\centerline {\small \it 12 Drahomanov St., Lviv, 79005, Ukraine}

\abstract{We find stringent upper bound on the momentum scale in noncommutative phase space of canonical type on the basis of studies of perihelion shift of the Mercury planet with taking into account features of description of motion of macroscopic body in the space with noncommutativity of coordinates and noncommutativity of momenta.  Using results for  precession of perihelion of the Mercury planet from ranging to the MESSENGER spacecraft we obtain upper bound for parameter of momentum noncommutativity  $10^{-80}\textrm{kg}^2\textrm{m}^2/\textrm{s}^2$ which is many orders less than known in the literature.

}

\section{Introduction}

Recently idea to describe features of space structure at the Planck scale considering modifications of commutation relations for coordinates and momenta has attracted much attention.
In the noncommutative phase space of canonical type relations for operators of coordinates and operators of momenta are as follows
  \begin{eqnarray}
[X_{i},X_{j}]=i\hbar\theta_{ij},\label{cform101}\\{}
[X_{i},P_{j}]=i\hbar(\delta_{ij}+\sigma_{ij}),\label{form1001}\\{}
[P_{i},P_{j}]=i\hbar\eta_{ij},\label{cform10001}{}
\end{eqnarray}
with  $\theta_{ij}$, $\eta_{ij}$, $\sigma_{ij}$ being elements of constant matrixes (see, for instance, \cite{Bertolami11,Bertolami06}). In the classical limit from (\ref{cform101})-(\ref{cform10001}) one obtains the corresponding Poisson brackets
  \begin{eqnarray}
\{X_{i},X_{j}\}=\theta_{ij},\label{form101}\\{}
\{X_{i},P_{j}\}=\delta_{ij}+\sigma_{ij},\label{form1001}\\{}
\{P_{i},P_{j}\}=\eta_{ij}.\label{form10001}{}
\end{eqnarray}

Various quantum and classical problems were studied in noncommutative space. Among them, for example, are free particles \cite{Djemai1,BastosPhysA,Laba,Harko}, hydrogen atom \cite{Chaichian,Ho,Chaichian1,Chair,Stern,Djemai1,Zaim2,Bertolamih,Adorno,Khodja,Alavi,GnatenkoPLA14,GnatenkoConf}, classical systems with various potentials \cite{Gamboa,Romero,Mirza,Djemai2,Daszkiewicz11,GnatenkoPLA13,GnatenkoPLA17,Daszkiewicz19} and many others. The studies are important for finding influence of space quantization on the properties of physical systems and for estimating the value of minimal length.

Extremely strong upper bounds on the minimal length in quantum space were obtained on the basis of studies of perihelion shift of the Mercury planet. Such studies were done in the frame of algebra with canonical noncommutativity of coordinates \cite{Romero,Mirza}, algebra with noncommutativity of coordinates and noncommutativity of momenta of canonical type \cite{Djemai2}, deformed algebra with minimal length \cite{Benczik02,Silagadze}, Snyder algebra  \cite{Ivetic}. For instance, in \cite{Benczik02} the upper bound for the minimal length which is $33$ orders less than the Planck length was obtained in deformed space.

In papers \cite{Tk2,GnatenkoPLA13,GnatenkoEPL19} it was concluded that the extremely small results for the minimal length obtained in \cite{Benczik02,Romero,Mirza,Djemai2,Ivetic} can be reexamined to more relevant one, taking into consideration features of description of motion of a macroscopic body in quantum space, namely taking into account that the motion of the center-of-mass of macroscopic body is described by effective parameters which are less than parameters corresponding to the elementary particles.
In this paper we show that taking into consideration features of description of motion of a macroscopic body in noncommutative phase space quite strong upper bound on the momentum scale  in the space can be obtained on the basis of studies of influence of noncommutativity on the perihelion shift of the Mercury planet.

Studies of a particle with mass $m$ in the gravitational field $-k/X$ ($k$ is a constant, $X=\sqrt{\sum_iX_i^2}$) in  noncommutative phase space of canonical type (\ref{form101})-(\ref{form10001}) were done in \cite{Djemai2}. Examining planar motion of the particle and considering $\sigma_{ij}=\sum_k\theta_{ik}\eta_{jk}/4$, $\theta_1=\theta_2=\eta_1=\eta_2=0$, $\theta=\theta_3$, $\eta=\eta_3$ with $\theta_i=\epsilon_{ijk}\theta_{jk}/2$, $\eta_i=\epsilon_{ijk}\eta_{jk}/2$,
 up to the first order in the parameters of noncommutativity the following expression for the perihelion shift of its orbit was obtained
 \begin{eqnarray}
\Delta\phi_{nc}=2\pi\left(\sqrt{\frac{m^2 k}{a^3(1-e^2)^3}}\theta+\frac{2}{e^2}\sqrt{\frac{a^3(1-e^2)^3}{m^2 k}}\eta\right),\label{shift}
\end{eqnarray}
where $a$, $e$ are the semi-major axis and eccentricity. The result was applied to the case of Mercury planet, substituting its mass, parameters of orbit into expression (\ref{shift}). On the basis of analysis of the values of multipliers $\sqrt{m k}/\sqrt{a^3(1-e^2)^3}$, and $2\sqrt{a^3(1-e^2)^3}/e^2\sqrt{m k}$, with $k=Gm_S$ ($G$ is the gravitational constant, $m_S$ is the mass of the Sun)  the contribution of the second term in (\ref{shift}) was ignored because of its smallness. Comparing result (\ref{shift}) with observed perihelion shift for the  Mercury planet the upper bound for the minimal length $\sqrt{\hbar\theta}\leq6.3\cdot10^{-33}$m  which is close to the Planck length was obtained \cite{Djemai2}.

In the present paper we study influence of noncommutativity on the perihelion shift of the Mercury planet, taking into consideration features of description of motion of macroscopic body in noncommutative phase space. Namely we take into account that the motion of composite system (macroscopic body) is described by effective parameters of noncommutativity.
 On the basis of results for precession of Mercury's perihelion from ranging to the MESSENGER spacecraft \cite{Park} we obtain stringent upper bound on the momentum scale in noncommutative phase space.

The paper is organized as follows.  In Section 2  composite system in noncommutative phase space of canonical type  (\ref{form101})-(\ref{form10001}) is considered. Features of motion of  free particle system and motion of a composite system in gravitational field are examined. We present conditions on the parameters of noncommutative algebra on which the motion of the center-of-mass of composite system is independent of the relative motion, a system of free particles with the same initial velocities does not fly away, and the weak equivalence principle is recovered in noncommutative phase space.  In Section 3 we examine influence of noncommutativity of coordinates and noncommutativity of momenta on the perihelion shift of the Mercury planet, taking into consideration features of description of macroscopic body motion in noncommutative phase space. The upper bounds on the parameters of noncommutativity are found. Conclusions are presented in Section 4.

\section{Motion of a composite system in a space with noncommutaivity of coordinates and noncommutativity of momenta}

Composite system in a space with noncommutativity of coordinates was studied in \cite{Ho,Bellucci,Daszkiewicz11,GnatenkoPLA13,Daszkiewicz13,GnatenkoU16,Daszkiewicz18}.
In a space with noncommutativity of coordinates and noncommutativity of momenta a two-particle problem was examined in  \cite{Djemai1}.
Features of motion of many-particle system were studied in four-dimensional (2D configurational and 2D momentum space) noncommutative phase space (\ref{form101})-(\ref{form10001}) with $i,j=1,2$ and $\gamma_{ij}=0$ \cite{GnatenkoPLA17,Laba}, in rotationally-invariant noncommutative phase space  constructed with the help of generalization of parameters of noncommutativity to tensors \cite{GnatenkoIJMPA18}.

Let us discuss features of description of a composite system in six-dimensional (3D configurational and 3D momentum space) noncommutative phase space of canonical type characterized by relations (\ref{form101})-(\ref{form10001}).
 Noncommutative algebra for coordinates and momenta of different particles can be written in the following form
  \begin{eqnarray}
\{X^{(n)}_{i},X^{(m)}_{j}\}=\delta_{nm}\theta^{(n)}_{ij},\label{m101}\\{}
\{X^{(n)}_{i},P^{(m)}_{j}\}=\delta_{nm}\delta_{ij}+\delta_{nm}\sigma^{(n)}_{ij},\label{m1001}\\{}
\{P^{(n)}_{i},P^{(m)}_{j}\}=\delta_{nm}\eta^{(n)}_{ij},\label{m10001}{}
\end{eqnarray}
here indexes $n$, $m$ label the particles. Parameters $\theta^{(n)}_{ij}$, $\eta^{(n)}_{ij}$, $\sigma^{(n)}_{ij}$
are considered to be different for different particles.
 Coordinates and momenta satisfying (\ref{m101}), (\ref{m10001}) can be represented by coordinates and momenta $x^{(n)}_i$, $p^{(n)}_i$ which satisfy the ordinary relations $\{x^{(n)}_{i},x^{(m)}_{j}\}=\{p^{(n)}_{i},p^{(m)}_{j}\}=0$, $\{x^{(n)}_{i},p^{(m)}_{j}\}=\delta_{mn}\delta_{ij}$ as
\begin{eqnarray}
X^{(n)}_i=x^{(n)}_i-\frac{1}{2}\sum_j{\theta}^{(n)}_{ij}p_j^{(n)},\label{sr}\\
P^{(n)}_i=p^{(n)}_i+\frac{1}{2}\sum_j{\eta}^{(n)}_{ij}x_j^{(n)}.\label{sr1}
\end{eqnarray}
Calculating Poisson brackets $\{X^{(n)}_{i},P^{(m)}_{j}\}$, one obtains
\begin{eqnarray}
\sigma^{(n)}_{ij}=\sum_k\frac{\theta^{(n)}_{ik}\eta^{(n)}_{jk}}{4},\label{ss}
\end{eqnarray}
(see \cite{Bertolami11,Bertolami06,Djemai2}).

For coordinates and momenta of the center-of-mass of a composite system (a macroscopic body) made of $N$ particles of masses $m_n$, taking into account (\ref{m101})-(\ref{m10001}), one has
  \begin{eqnarray}
\{X^{c}_{i},X^{c}_{j}\}=\theta^{c}_{ij},\label{r101}\\{}
\{X^{c}_{i},P^{c}_{j}\}=\delta_{ij}+\sum_n\mu_n\sigma^{(n)}_{ij},\label{r1001}\\{}
\{P^{c}_{i},P^{c}_{j}\}=\eta^{c}_{ij},\label{r10001}{}
\end{eqnarray}
where $X_i^c=\sum_{n}\mu_{n}X_i^{(n)}$, ${P}_i^c=\sum_{n}{P}_i^{(n)}$, $\mu_n=m_n/M$, $M=\sum_nm_n$, parameters  $ \theta^{c}_{ij}$, and  $\eta^{c}_{ij}$ are defined as
\begin{eqnarray}
\theta^{c}_{ij}=\sum_n\mu_n^2 \theta^{(n)}_{ij},\label{eft}\\
\eta^{c}_{ij}=\sum_n \eta^{(n)}_{ij}. \label{eft1}
\end{eqnarray}
Note that relations for coordinates and momenta of the center-of-mass of composite system are not the same as relations of noncommutative algebra for particles forming the system (\ref{m101})-(\ref{m10001}).
Poisson brackets for coordinates of the center-of-mass and Poisson brackets for momenta of the center-of-mass are equal to the effective parameters of noncommutativity (\ref{eft}), (\ref{eft1}) and in  expression (\ref{r1001}) one has $\sum_n\mu_n\sigma^{(n)}_{ij}=\sum_n\mu_n\sum_k{\theta^{(n)}_{ik}\eta^{(n)}_{ik}}/{4}\neq\sum_k{\theta^{c}_{ik}\eta^{c}_{jk}}/{4}$.

In addition it is worth mentioning that in noncommutative phase space of canonical type the motion of the center-of-mass is not independent of the relative motion. For coordinates and momenta of the center-of-mass and coordinates and momenta of the relative motion  defined in the traditional way  one has
\begin{eqnarray}
\{X^c_i,\Delta X^{(n)}_j\}=\mu_n \theta^{(n)}_{ij}-\sum_m\mu_m^2\theta^{(m)}_{ij},\label{nez1}\\
\{P^c_i,\Delta P^{(n)}_j\}=\eta^{(n)}_{ij}-\mu_n\sum_m\eta^{(m)}_{ij},\\
\{\Delta X^{(n)}_i, P^{(c)}_j\}=\sigma^{(n)}_{ij}-\sum_m\mu_m\sigma^{(m)}_{ij},\\
\{X^{c}_i, \Delta P^{(n)}_j\}=\mu_n(\sigma^{(n)}_{ij}-\sum_m\mu_m\sigma^{(m)}_{ij}),\label{nez2}
\end{eqnarray}
where $\Delta X^{(n)}_i=X^{(n)}_i-X^{c}_i$, $\Delta P^{(n)}_i=P^{(n)}_i-\mu_n P^{c}_i$.

Let us consider the following conditions on the parameters $\theta_{ij}^{(n)}$, $\eta_{ij}^{(n)}$
\begin{eqnarray}
\theta_{ij}^{(n)}m_n=\gamma_{ij}, \label{cco1}\\
\frac{\eta_{ij}^{(n)}}{m_n}=\alpha_{ij}, \label{ccoo1}
\end{eqnarray}
here $\gamma_{ij}$, $\alpha_{ij}$ are constants which do not depend on mass.
From (\ref{ss}), (\ref{cco1}), (\ref{ccoo1}) we have that parameters $\sigma_{ij}^{(n)}$ are the same for different particles
\begin{eqnarray}
\sigma^{(n)}_{ij}=\sum_k\frac{\gamma_{ik}\alpha_{jk}}{4}=\sigma_{ij}.\label{o}
\end{eqnarray}
We would like to stress that if conditions (\ref{cco1})-(\ref{o}) are satisfied, namely if parameters of coordinate noncommutativity are proportional inversely to mass, parameters of momentum noncommutativity are proportional to mass and therefore parameters $\sigma^{(n)}_{ij}$ are the same for particles with different masses, the relations (\ref{nez1})-(\ref{nez2}) have the form
\begin{eqnarray}
\{X^c_i,\Delta X^{(n)}_j\}=\{P^c_i,\Delta P^{(n)}_j\}=0,\\
\{\Delta X^{(n)}_i, P^{(c)}_j\}=\{X^{c}_i, \Delta P^{(n)}_j\}=0.
\end{eqnarray}
So, in this case the motion of the center-of-mass of a body can be considered independently of the relative motion.

In addition, taking into account (\ref{ss}), (\ref{eft}), (\ref{eft1}), (\ref{cco1})-(\ref{o}),  one has
\begin{eqnarray}
\sigma_{ij}=\sum_k\frac{\theta^{c}_{ik}\eta^{c}_{jk}}{4}=\sum_k\frac{\theta^{(n)}_{ik}\eta^{(n)}_{jk}}{4}.\label{sigm1}
\end{eqnarray}
So, the relations for coordinates of the center-of-mass reproduce relations of noncommutative algebra for coordinates and momenta of particles (\ref{m101})-(\ref{m10001}) with effective parameters of noncommutativity
\begin{eqnarray}
\theta^{c}_{ij}=\frac{\gamma_{ij}}{M},\label{eft2}\\
\eta^{c}_{ij}= M\alpha_{ij}.\label{eft22}
\end{eqnarray}
Note, that the effective parameters of noncommutativity which correspond to composite system do not depend on its composition and are determined by its mass $M$ (\ref{eft2}), (\ref{eft22}), similarly as parameters of noncommutativity corresponding to the individual particles  are determined by their masses (\ref{cco1}), (\ref{ccoo1}).

 In addition we would like to mention that on the conditions (\ref{cco1})-(\ref{o}) the weak equivalence principle is recovered in noncommutative phase space.
 Note that expression for the perihelion shift (\ref{shift}) depends on mass \cite{Djemai2}. It is a consequence of violation of the equivalence principle in noncommutative phase space of canonical type.
 Effect of noncommutativity on the implementation of the equivalence principle was studied in \cite{GnatenkoPLA13,Saha,Saha1,GnatenkoPRD,GnatenkoMPLA19,Bastos11,GnatenkoEPL18,Bertolami15}. In \cite{Bertolami15} it was concluded that the equivalence principle holds in the sense that an accelerated frame of reference is locally equivalent to a gravitational field, unless noncommutative parameters are anisotropic, $\eta_{xy}\neq\eta_{yz}$. In \cite{GnatenkoPLA13,GnatenkoPLA17,GnatenkoEPL18} it was shown that the weak equivalence principle can be recovered in noncommutative space of canonical type, in four-dimensional noncommutative phase space, in rotationally-invariant noncommutative phase space, considering parameters of noncommutativity to be dependent on mass. This conclusion can be generalized to the case of algebra (\ref{form101})-(\ref{form10001}).
  If relations (\ref{cco1}), (\ref{ccoo1}) hold the trajectory of a particle (a body) in gravitational field $V({\bf{X}})$ does not depend on its mass and composition.
 For a particle with mass $m$, considering Hamiltonian
 \begin{eqnarray}
 H=\frac{P^2}{2m}+m V(\bf{X}),
 \end{eqnarray}
  and taking into account relations (\ref{form101})-(\ref{form10001}) one finds
  \begin{eqnarray}
\dot{X}_i=\sum_j(\delta_{ij}+\sigma_{ij})\frac{P_j}{m}+\sum_jm\theta_{ij}\frac{\partial V}{\partial X_j},\label{mmm1}\\
\dot P_i=-\sum_j(\delta_{ij}+\sigma_{ij})\frac{\partial V}{\partial X_j}+\sum_j\eta_{ij}\frac{P_j}{m}.\label{mmm2}
\end{eqnarray}
 Note that if relations (\ref{cco1})-(\ref{o}) hold one can rewrite (\ref{mmm1}), (\ref{mmm2}) as
  \begin{eqnarray}
\dot{X}_i=\sum_j(\delta_{ij}+\sigma_{ij})P^{\prime}_j+\sum_j\gamma_{ij}\frac{\partial V}{\partial X_j},\label{mm1}\\
\dot P^{\prime}_i=-\sum_j(\delta_{ij}+\sigma_{ij})\frac{\partial V}{\partial X_j}+\sum_j\alpha_{ij}{P^{\prime}_j}.\label{mm2}
\end{eqnarray}
Equations (\ref{mm1}), (\ref{mm2}) do not depend on mass, therefore $X_i(t)$ and $P^{\prime}_i(t)$ ($P^{\prime}_i=P_i/m$) do not depend on mass too. So, the weak equivalence principle which states that the motion of a particle in gravitational field is independent of its mass and composition is preserved.

 In the case of macroscopic body, if relations (\ref{cco1})-(\ref{o}) are satisfied the motion of the center-of-mass of a body in noncommutative phase space can be studied independently of the relative motion, the coordinates and the momenta of the center-of-mass satisfy noncommutative algebra with effective parameters of noncommutativity (\ref{eft2}), (\ref{eft22}) which do not depend on the composition of the body. Therefore, for macroscopic body in gravitational field the equations of motion have the form (\ref{mmm1}), (\ref{mmm2}) with parameters (\ref{eft2}), (\ref{eft22}) and can be rewritten as (\ref{mm1}), (\ref{mm2}). So, the motion of a body in gravitational field in noncommutative phase space does not depend on its mass and composition and the weak equivalence principle is satisfied.

Note also that if conditions  (\ref{cco1}), (\ref{ccoo1}) hold, namely if
$\theta m =\gamma$,  $\eta/{m}=\alpha$ ($\gamma$, $\alpha$ are constants which do not depend on mass) expression for perihelion shift (\ref{shift}) can be rewritten as
\begin{eqnarray}
\Delta\phi_{nc}=2\pi\left(\sqrt{\frac{k}{a^3(1-e^2)^3}}\gamma+\frac{2}{e^2}\sqrt{\frac{a^3(1-e^2)^3}{k}}\alpha\right),\label{shiftt}
\end{eqnarray}
 and does not depend on mass.

 Besides it is worth noting that due to relations (\ref{cco1})-(\ref{o}) the motion of free particle in noncommutative phase space does not depend on its mass.  The equations of motion of free particle with mass $m$ in the space (\ref{form101})-(\ref{form10001})  have the form
\begin{eqnarray}
\dot{X}_i=\sum_j(\delta_{ij}+\sigma_{ij})\frac{P_j}{m},\label{m1}\\
\dot P_i=\sum_j\eta_{ij}\frac{P_j}{m}.\label{m2}
\end{eqnarray}
From the equations one finds
\begin{eqnarray}
\dot{X}_i(t)=A_{i1}\cos\left(\frac{\tilde{\eta}}{m}t\right)+A_{i2}\sin\left(\frac{\tilde{\eta}}{m}t\right)+A_{i3},\label{v}\\
\tilde{\eta}=\sqrt{\eta^2_{12}+\eta^2_{23}+\eta^2_{31}},\label{te}
\end{eqnarray}
where $A_{ij}$ are elements of matrix
\begin{eqnarray}
\hat{A}=(1+\hat{\sigma})\times\nonumber\\
\times\quad
\begin{pmatrix}
\frac{C_2\eta_{31}\tilde{\eta}-C_1\eta_{12}\eta_{23}}{\eta^2_{23}+\eta^2_{31}} &-\frac{C_1\eta_{31}\tilde{\eta}+C_2\eta_{12}\eta_{23}}{\eta^2_{23}+\eta^2_{31}} & \frac{C_3\eta_{23}}{\eta_{12}} \\
-\frac{C_2\eta_{23}\tilde{\eta}+C_1\eta_{12}\eta_{31}}{\eta^2_{23}+\eta^2_{31}} & \frac{C_1\eta_{23}\tilde{\eta}-C_2\eta_{12}\eta_{31}}{\eta^2_{23}+\eta^2_{31}} & \frac{C_3\eta_{23}}{\eta_{12}} \\
C_1  & C_2  & C_3
\end{pmatrix}
\quad
\end{eqnarray}
Here constants $C_i$ are determined by the initial velocities $\upsilon_{0i}$
\begin{eqnarray}
(1+\hat{\sigma})\hat{B}\hat{C}=
\hat{\upsilon}_0,
\end{eqnarray}
where
\begin{eqnarray}
\hat{B}=
\quad
\begin{pmatrix}
\frac{-\eta_{12}\eta_{23}}{\eta^2_{23}+\eta^2_{31}} & \frac{\eta_{31}\tilde{\eta}}{\eta^2_{23}+\eta^2_{31}} & \frac{\eta_{23}}{\eta_{12}} \\
\frac{-\eta_{12}\eta_{31}}{\eta^2_{23}+\eta^2_{31}} & -\frac{\eta_{23}\tilde{\eta}}{\eta^2_{23}+\eta^2_{31}} & \frac{\eta_{31}}{\eta_{12}} \\
1  & 0  & 1
\end{pmatrix}
\quad\\
\hat{C}=\quad
\begin{pmatrix}
C_1\\
C_2\\
C_3
\end{pmatrix}
\quad \ \   \hat{\upsilon}_0=\quad
\begin{pmatrix}
\upsilon_{01}\\
\upsilon_{02}\\
\upsilon_{03}
\end{pmatrix}
\quad
\end{eqnarray}
Matrix $\hat{\sigma}$ has elements $\sigma_{ij}$ (\ref{ss}).

We would like to stress that the velocity of free particle (\ref{v}) and therefore its trajectory depend on its mass. So, because of noncommutativity of momenta free particles with the same initial velocities but different masses fly away. For a system of free particles with the same initial velocities, the velocity of the center-of-mass is not equal to the velocities of particles forming the system, the relative velocities of the particles are not equal to zero.

Note that if relations (\ref{cco1})-(\ref{o}) are satisfied for a free particle one can write equations (\ref{mm1}), (\ref{mm2}) with $V=0$. Solutions of the equations $X_i(t)$, $P_i^{\prime}(t)$ do not depend on mass. Therefore if conditions (\ref{cco1})-(\ref{o}) are satisfied for a system of free particles with the same initial velocities, taking into account (\ref{v}), one has that the velocity of the center-of-mass of free particle system is equal to the velocities of particles forming it
\begin{eqnarray}
\dot{X}^{c}_i(t)=\sum_n \mu_n \dot{X}^{(n)}_i(t)=\sum_{n}\mu_n\left(A^{(n)}_{i1}\cos\left(\frac{\tilde{\eta}^{(n)}}{m_n}t\right)+\right.\nonumber\\ \left.+A^{(n)}_{i2}\sin\left(\frac{\tilde{\eta}^{(n)}}{m_n}t\right)+A^{(n)}_{i3}\right)=\nonumber\\=A_{i1}\cos\left(\sqrt{\alpha^2_{12}+\alpha^2_{23}+\alpha^2_{31}}t\right)+\nonumber\\+A_{i2}\sin\left(\sqrt{\alpha^2_{12}+\alpha^2_{23}+\alpha^2_{31}}t\right)+A_{i3}=\dot{X}_i^{(n)}(t),\label{v2}
\end{eqnarray}
and relative velocities are equal to zero
\begin{eqnarray}
 \Delta\dot{X}_i(t)=\dot{X}^{(n)}_i(t)-\dot{X}^{c}_i(t)=0,
 \end{eqnarray}
 as it is in the ordinary space ($\theta_{ij}=\eta_{ij}=0$).
  Writing  (\ref{v2}) we take into account that due to relation (\ref{ccoo1}) we can write
  \begin{eqnarray}
\frac{\tilde{\eta}^{(n)}}{m_n}=\frac{\sqrt{(\eta^{(n)}_{12})^2+(\eta^{(n)}_{23})^2+(\eta^{(n)}_{31})^2}}{m_n}=\nonumber\\=\sqrt{\alpha^2_{12}+\alpha^2_{23}+\alpha^2_{31}}.
\end{eqnarray}
So, arguments of sine and cosine do not depend on mass. Also if relation (\ref{ccoo1}) holds elements of matrix $\hat{A}$ depend on the constants $\alpha_{ij}$ which are the same for different particles and do not depend on masses. Therefore, one can write $A^{(n)}_{ij}=A_{ij}$.

So, if conditions (\ref{cco1})-(\ref{o}) are satisfied  the motion of the center-of-mass of a composite system is independent of the relative motion; coordinates and momenta of the center-of-mass satisfy noncommutative algebra with parameters of noncommutativity which are independent of its composition; trajectory of free particle does not depend on its mass; a system of free particles with the same initial velocities does not fly away; the weak equivalence principle is recovered in noncommutative phase space (\ref{form101})-(\ref{form10001}).

It is worth adding that the  idea to relate parameters of deformed algebra with mass opens possibility to obtain  important results in deformed space with minimal length \cite{Tk3},  in noncommutative space of canonical type \cite{GnatenkoPLA17},  in twist-deformed space \cite{GnatenkoMPLA19}, in a space with Lie-algebraic noncommutativity \cite{GnatenkoPRD}.

\section{Estimation of upper bounds on the parameters of noncommutativity on the basis of studies of precession of Mercury's perihelion}

Let us study influence of noncommutativity of coordinates and noncommutativity of momenta on the perihelion shift of the Mercury planet with taking into account features of description of motion of  macroscopic body in noncommutative phase space.

As was shown in the previous section motion of macroscopic body in noncommutative phase space is described by effective parameters of noncommutativity (\ref{eft}), (\ref{eft1}). So,  parameters $\theta$, $\eta$   in (\ref{shift}) which are determined as $\theta=\theta_3=\theta_{12}$, $\eta=\eta_3=\eta_{12}$ (see \cite{Djemai2}) should be replaced by $\theta^{c}=\sum_n\mu_n^2\theta^{(n)}$, $\eta^{c}=\sum_n \eta^{(n)}$, where $\theta^{(n)}$, $\eta^{(n)}$ are parameters of noncommutativity corresponding to particles with masses $m_n$ which form the planet ($\theta^{(n)}=\theta^{(n)}_3=\theta^{(n)}_{12}$, $\eta^{(n)}=\eta^{(n)}_3=\eta^{(n)}_{12}$). So, we have
\begin{eqnarray}
\Delta\phi_{nc}=\Delta\phi_{\theta}+\Delta\phi_{\eta},\label{shiftM}\\
\Delta\phi_{\theta}=2\pi\sqrt{\frac{G m^2_M m_S}{a^3(1-e^2)^3}}\theta^c,\\
\Delta\phi_{\eta}=\frac{4\pi}{e^2}\sqrt{\frac{a^3(1-e^2)^3}{G m^2_M m_S}}\eta^{c}.
\end{eqnarray}

Writing \label{shiftM} we assume that influence of relative motion on the motion of the center-of-mass of the Mercury planet can be neglected and take into account that $k=Gm_S$, $G$ is the gravitational constant and $m_S$ is the mass of the Sun.

Observed perihelion precession rate which cannot be explained by the Newtonian gravitational effects of other planets and asteroids, Solar Oblateness  is  \cite{Park}
\begin{eqnarray}
\Delta\phi_{obs}=42.9779\pm0.0009\textrm{ arc-seconds per century}=\nonumber\\=2\pi(7.98695\pm0.00017)\cdot10^{-8}\textrm{radians/revolution}.
 \end{eqnarray}
Assuming that the perihelion shift of the Mercury planet caused by noncommutativity of coordinates and noncommutativity of momenta is less than $\Delta\phi_{obs}$ one can write
$|\Delta\phi_{\theta}+\Delta\phi_{\eta}|\leq |\Delta\phi_{obs}|$.
To estimate the orders of parameters of noncommutativity, it is sufficiently to consider the following inequalities
$|\Delta\phi_{\theta}|\leq |\Delta\phi_{obs}|$, $|\Delta\phi_{\eta}|\leq|\Delta\phi_{obs}|$,
from which one obtains
\begin{eqnarray}
\hbar|\theta^{c}|\leq 10^{-59}\textrm{m}^2,\label{estt}\\
\hbar|\eta^{c}|\leq 10^{-26}\textrm{kg}^2\textrm{m}^2/\textrm{s}^2.\label{este}
\end{eqnarray}
Let us find effective parameters of noncommutativity $\theta^{c}$, $\eta^{c}$ corresponding to the Mercury planet.
Taking into account (\ref{eft1}) for effective parameter of momentum noncommutativity  one has
\begin{eqnarray}
\eta^c=N_{nuc}\eta^{(nuc)}+N_e\eta^{(e)},\label{eM}
\end{eqnarray}
here $N_{nuc}$ is the number of nucleons and $N_{e}$ is the  number of electrons in the planet, $\eta^{(nuc)}$, $\eta^{(e)}$ are parameters of noncommutativity corresponding to nucleons and electrons, respectively.  The main contribution to the mass of the planet comes from nucleons. Therefore, their number can be calculated as $N_{nuc}\simeq m_M/m_{nuc}$, $m_{nuc}$ is the mass of nucleon.  Taking into account that the number of electrons in the planet is equal to the number of protons $N_{p}$  and $N_{p}\simeq N_{nuc}/2$, one has $N_e\simeq N_{nuc}/2$. The nucleons are made of three quarks, so for the parameter of noncommutativity corresponding to nucleons we can write $\eta^{(nuc)}=3\eta^{(q)}$ ($\eta^{(q)}$ is parameter of momentum noncommutativity corresponding to quark). As a result, assuming that parameters of noncommutativity corresponding to the elementary particles (electrons and quarks) are of the same order, on the basis of (\ref{eM}) one obtains
\begin{eqnarray}
\eta^{c}\simeq3N_{nuc}\eta^{(q)}+\frac{N_{nuc} \eta^{(e)}}{2}\simeq \frac{m_{M}}{m_{nuc}}\eta^{(nuc)}.\label{effeq}
 \end{eqnarray}
So, on the basis of (\ref{este}), for parameter of momentum noncommutativity, corresponding to nucleons, we find
\begin{eqnarray}
\hbar|\eta^{(nuc)}|\leq 10^{-76}\textrm{kg}^2\textrm{m}^2/\textrm{s}^2.\label{umn1}
\end{eqnarray}

 Analogically, taking into account expression for the effective parameter of coordinate noncommutativity (\ref{eft}) for the Mercury planet one has
\begin{eqnarray}
\theta^c=N_{nuc}\theta^{(nuc)}\frac{m^2_{nuc}}{m^2_M}+N_{e}\theta^{(e)}\frac{m^2_{e}}{m^2_M}\simeq\frac{\theta^{(nuc)}m_{nuc}}{m_M},\label{efftq}
\end{eqnarray}
here $\theta^{(nuc)}$, $\theta^{(e)}$ are parameters of noncommutativity corresponding to nucleons and electrons, $m_e$ is the mass of electron. The details of calculation of the effective parameter of coordinate noncommutativity for the Mercury planet can be found in \cite{GnatenkoPLA13}.
Therefore, from (\ref{estt}) one obtains
\begin{eqnarray}
\hbar|\theta^{(nuc)}|\leq 10^{-9}\textrm{m}^2.\label{ut1}
\end{eqnarray}

Note, that the same results for the upper bounds for parameters of noncommutativity  (\ref{umn1}), (\ref{ut1}) can be obtained considering relations (\ref{cco1}), (\ref{ccoo1}). If conditions (\ref{cco1}), (\ref{ccoo1}) hold taking into account (\ref{eft2}), (\ref{eft22}) we obtain $\eta^{(nuc)}/m_{nuc}=\eta^c/m_M$, $\theta^{(nuc)}m_{nuc}=\theta^c m_M$  that correspond  to (\ref{effeq}), (\ref{efftq}).

From General Relativity predictions the perihelion
precession rate is $\Delta\phi_{GR}=2\pi(7.98744 \cdot 10^{-8})$radians/revolution (see, for instance, \cite{Benczik02}).
Comparing the perihelion shift caused by noncommutativity  with
 \begin{eqnarray}
 \Delta\phi_{obs}-\Delta\phi_{GR}=\nonumber\\=2\pi (-0.00049\pm 0.00017)\cdot 10^{-8}\textrm{radians/revolution},
  \end{eqnarray}
and assuming that $|\Delta\phi_{nc}|$ is less than $|\Delta\phi_{obs}-\Delta\phi_{GR}|$ at $3\sigma$ one can write
\begin{eqnarray}
|\Delta\phi_{nc}|\leq 2\pi\cdot10^{-11}\textrm{radians/revolution}.\label{eq}
\end{eqnarray}
Similar assumption was considered in  \cite{Benczik02} for estimation of the minimal length in the deformed space, and in \cite{Romero,Mirza,Djemai2} for estimation of the minimal length in noncommutative space of canonical type.
Considering
\begin{eqnarray}
|\Delta\phi_{\theta}|\leq 2\pi\cdot10^{-11}\textrm{radians/revolution},\\
|\Delta\phi_{\eta}|\leq 2\pi\cdot10^{-11}\textrm{radians/revolution},
\end{eqnarray}
 we find
\begin{eqnarray}
\hbar|\theta^{(nuc)}|\leq 7.2\cdot10^{-13}\textrm{m}^2,\label{umc1}\\
\hbar|\eta^{(nuc)}|\leq 3.3\cdot10^{-80}\textrm{kg}^2\textrm{m}^2/\textrm{s}^2.\label{umn2}
\end{eqnarray}

The upper bound for the parameter of coordinate noncommutativity (\ref{umc1})  is in agreement with the result obtained on the basis of studies of neutrons in gravitational well in noncommutative space \cite{Castello-Branco}.
Note that the upper bounds for parameter of momentum noncommutativity (\ref{umn1}), (\ref{umn2}) are quite strong. The result (\ref{umn2}) is 13 orders less than that obtained on the basis of studies of neutrons in gravitational quantum well in noncommutative phase space \cite{Bertolami05} and 14 orders less than the upper bound obtained examining effect of noncommutativity on the hyperfine structure of hydrogen atom in noncommutative phase space \cite{Bertolamih}.

 From  (\ref{umn2}) we obtain that the upper bound on the momentum scale is
\begin{eqnarray}
\sqrt{\hbar|\eta^{(nuc)}|}\leq 1.8\cdot10^{-40}\textrm{kg}\cdot\textrm{m}/\textrm{s}.\label{9}
  \end{eqnarray}
To analyze the obtained result it is worth to compare it with known values. From the Heisenberg uncertainty relation one has $\Delta P\geq\hbar/2\Delta X$. For the distance corresponding to diameter of the observable universe $8.8\cdot10^{26}$m \cite{Bars} from this relation we find  $\Delta P\geq6\cdot10^{-62}\textrm{kg}\cdot\textrm{m}/\textrm{s}$.  The obtained upper bound (\ref{9}) is far from this value, one has $\sqrt{\hbar|\eta^{(nuc)}|}/\Delta P\simeq10^{21}$.

\section{Conclusion}

A space with noncommutativity of coordinates and noncommutativity of momenta of canonical type (\ref{form101})-(\ref{form10001}) has been considered. Features of description of motion of a composite system in the space have been discussed. It has been shown that  if parameters of noncommutative algebra are related with mass, namely if conditions (\ref{cco1})-(\ref{o}) are satisfied, the  motion of the center-of-mass of composite system  is independent of the relative motion, the system of free particles with the same initial velocities does not fly away, the weak equivalence principle is recovered in the noncommutative phase space  (\ref{form101})-(\ref{form10001})

Perihelion shift of the Mercury planet  has been studied with taking into consideration features of description of motion of macroscopic body in noncommutative phase space. On the basis of results of these studies and results for precession of Mercury's perihelion from ranging to the MESSENGER spacecraft we have estimated the upper bound on the parameters of noncommutativity corresponding to nucleons.
We have concluded that taking into account expressions for effective parameters of noncommutativity (\ref{effeq}), (\ref{efftq}), corresponding to the Mercury planet,
extremely strong upper bound for the minimal length presented in \cite{Djemai2} can be reexamined to more relevant result (\ref{umc1})  and quite stringent upper bound on the parameter of momentum noncommutativity can be found.
 Namely we have found upper bound $10^{-80}\textrm{kg}^2\textrm{m}^2/\textrm{s}^2$  for the parameter of momentum noncommutativity corresponding to nucleons. This result is many orders less than that obtained on the basis of studies of neutrons in gravitational field and on the basis of studies of the hydrogen atom in noncommutative phase space  \cite{Bertolamih,Bertolami05}.

\section*{Acknowledgements}

The authors thank Prof. Novosyadlyj B. S. for useful comments.  This work was partly supported by the Project $\Phi\Phi$-63Hp
(No. 0117U007190) from the Ministry of Education and Science of Ukraine.

\end{document}